\documentclass{elsart}
\usepackage{epsfig}
\newcommand{\be}{\begin{eqnarray}}
\newcommand{\ee}{\end{eqnarray}}
\newcommand{\bra}[1]{\mbox{$\langle\, #1 \mid$}}

\newcommand{\ket}[1]{\mbox{$\mid #1\,\rangle$}}

\newcommand{\expec}[1]{\mbox{$\langle\, #1\,\rangle$}}

\renewcommand{\natural}{\mbox{{\rm I\hspace{-2truemm} N}}}

\begin{document}
\begin{frontmatter}
\title{On time evolution of quantum black holes}
\author[a]{Roberto Casadio}
\address[a]{Dipartimento di Fisica, Universit\`a di Bologna, and
I.N.F.N, Sezione di Bologna, via Irnerio 46, 40126 Bologna, Italy
\thanksref{e}}
\thanks[e]{Email: casadio@bo.infn.it}
\begin{abstract}
The time evolution of black holes involves both the canonical equations
of quantum gravity and the statistical mechanics of Hawking radiation,
neither of which contains a time variable.
In order to introduce the time, we apply the semiclassical approximation
to the Hamiltonian constraint on the apparent horizon and show that,
when the backreaction is included, it suggests the existence of a
long-living remnant, similarly to what is obtained in the microcanonical
picture for the Hawking radiation.
\end{abstract}
\end{frontmatter}
%
%
%
%\section{Introduction}
%\setcounter{equation}{0}
%
%
%
A longstanding issue in theoretical physics is how to quantize
Einstein gravity.
Since in the theory there are constraints whose algebra
contains structure functions (which depend on the space-time
coordinates), one can conceive (inequivalent \cite{hajicek})
manners of lifting the Hamiltonian and momentum constraints
to quantum equations, thus hindering the formal solution
to the problem.
On a more physical ground, one might notice that it is relevant
to have a quantum theory of gravity at our disposal only for
systems with very strong (Planck size) gravitational fields, such
as those one expects in the early stages of the Universe.
Because of Hawking's discovery of black hole evaporation
\cite{hawking}, one also expect that quantum (or semiclassical)
gravity plays a role in determining the dynamics of the late stages
in the life of collapsed objects.
\par
In a general situation, one has to deal with the infinite number of
degrees of freedom of the gravitational field configurations which
are physically distinguished only modulo coordinate transformations.
This gives the constraints the form of functional differential
equations, for which there is little hope of finding general
solutions, and one then tries to reduce the number of degrees of
freedom from the onset.
One way of implementing this program is the {\em background field
method\/} \cite{dewitt1,birrell}, in which one assumes a certain
background metric manifold and does perturbation theory on it
(although it turns out to be non-renormalizable).
Another method is to impose a large (space-time) symmetry
{\em prior\/} to quantization, resulting in a so called
{\em mini-} (or {\em midi-}) {\em superspace\/} for a finite
(or discrete) number of canonical variables \cite{dewitt,wheeler}.
In the latter approach to quantum gravity there is no time
and one expects that a time variable can be identified
by resorting to the semiclassical approximation \cite{bv}.
\par
Unfortunately, there are systems which do not allow for such
a reduction, and (spherical) black holes fall among these cases,
because one cannot find a finite set of variables in which to
express both the Hamiltonian and the momentum constraints
consistently over the whole space-time manifold \cite{lousto}.
Years ago, Tomimatsu \cite{tomimatsu} proposed an alternative
scheme of reduction by considering the support of the constraints
just near the apparent horizon and quantizing the residual
degrees of freedom.
\par
Hawking radiation can also be studied from the point of view
of the statistical mechanics by assuming the horizon area
${\mathcal A}$ is a measure of the internal (microscopical)
degeneracy $\sigma$ of black holes according to
\cite{bekenstein}~\footnote{This relation suggests that black
holes are extended $p$-branes \cite{micro} and was also obtained
in string theory \cite{strings}.}
\be
\sigma\sim \exp\left({{\mathcal A}\over 4}\right)
\ ,
\ee
and thus determines the probability
\be
P\simeq {c\over\sigma}
\ ,
\label{c}
\ee
for Hawking particles to tunnel out through the horizon.
When the space-time is asymptotically flat, it is known that the
canonical ensemble is formally inconsistent and one should instead
use the microcanonical ensemble in order to ensure global energy
conservation in the system \cite{micro}.
For large black holes the two descriptions are practically
equivalent, but they then disagree when the mass becomes small
(of order the Planck mass).
Within this approach the detailed form of the metric (including the
backreaction) affects the grey-body factor \cite{page} and
the constant $c$ in Eq.~(\ref{c}) cannot be determined without a complete
knowledge of the (non-local) physics near the horizon.
In this respect, the microcanonical picture is complementary to the
dynamical approach of Ref.~\cite{tomimatsu}.
\par
The aim of this letter is precisely to study the consequences which can be
derived from confronting the constraints of canonical (quantum) gravity
localized on the apparent horizon as in Ref.~\cite{tomimatsu} with the global
microcanonical picture of the Hawking effect, with a particular
emphasis on the role played by the backreaction of the emitted radiation.
\par
Let us start from the dynamical approach.
Tomimatsu \cite{tomimatsu} considered a conformally coupled scalar field
$\phi$ in a four-dimensional spherically symmetric space-time with action
\cite{thomi}
\be
S&=&{1\over 2}\,\int d^2x\,\sqrt{-g}\,\left[
1+g^{ab}\,\partial_a\varphi\,\partial_b\varphi
+{1\over 2}\,R\,\varphi^2
-\varphi^2\,g^{ab}\,\partial_a\phi\,\partial_b\phi
\right]
\ .
\label{action}
\ee
The four-dimensional metric has been written as
\be
ds^2=g_{ab}\,dx^a\,dx^b+\varphi^2\,d\Omega^2
\ ,
\ee
where $x^a=(t,r)$, $d\Omega^2$ is the line element of a unit
two-sphere, $\varphi=\varphi(t,r)$ the radius of a two-sphere
of area $4\,\pi\,\varphi^2$, $g_{ab}=g_{ab}(t,r)$ a two-dimensional
metric and $R$ its scalar curvature.
Adopting the ADM decomposition of a generic two-dimensional
metric \cite{mtw}
\be
g_{ab}=\left[
\begin{array}{cc}
-N^2+\strut\displaystyle{N_r^2\over\gamma}\ \ \ \ & N_r\\
N_r & \gamma
\end{array}\right]
\ ,
\ee
with $N$ the lapse function and $N_r$ the (radial) shift function,
one obtains that the momentum $H_r$ and the Hamiltonian $H$
coincide~\footnote{In a suitable gauge with $\gamma=2$ and its
conjugate momentum $\Pi_\gamma=\varphi/4$.
Consequently $\gamma$ is not dynamical \cite{tomimatsu}.}
near the apparent horizon (defined by $g_{tt}=0$),
$H_r=H/\sqrt{2}$.
This just leaves the Wheeler-DeWitt equation (Hamiltonian constraint)
for the wavefunction $\Psi=\Psi(\phi,m)$,
\be
{1\over 2}\,\left[{\hat\Pi_\phi^2\over 4\,m^2}
-2\,\hat\Pi_m+{1\over 2}\right]
\Psi=0
\ ,
\label{wdw}
\ee
where $m$ is the black hole mass (in units with $c=\hbar=G=1$)
as it appears in $g_{tt}=-1+2\,m/r$.
The classical momenta are given by
\be
\Pi_m=
{1\over 2}\,\left(\dot m+{1\over 2}\right)
\ , \ \ \
\Pi_\phi=4\,m^2\,\dot\phi
\ ,
\ee
and a dot denotes derivative with respect to the time $t$
measured in the reference frame of a (distant) static
observer.
To summarize, the system near the apparent horizon has two physical
degrees of freedom: the Bondi mass $m$ computed on the outer surface
of the apparent horizon (of radius $\varphi=2\,m$) and the scalar
field $\phi$ which can be used to reproduce the outgoing Hawking
radiation as we shall show below (see also Ref.~\cite{tomimatsu}).
\par
Eq.~(\ref{wdw}) can be solved exactly by separation of variables,
and one finds out the set of solutions
\be
\Psi(m,\phi)=\Psi_0\,\exp\left\{i\,\left(
{k^2\over 8\,m}-{m\over 4}-k\,\phi\right)\right\}
\ ,
\label{exact}
\ee
where $\Psi_0$ is a constant and $k$ a continuous (complex) wave
number.
For $k$ real one has ``plane waves'' corresponding to a
static black hole, and for $k$ imaginary Hawking's behaviour
\cite{hawking} is recovered \cite{tomimatsu}.
Despite the simple form of the derivation, the above solution
taken at face value is not very illuminating about the temporal
evolution of the black hole, since there is no time dependence in the
function in Eq.~(\ref{exact}), and one needs, at least in principle,
to build a suitable superposition of modes in such a way that a
time variable can be defined (in terms of $m$ and $\phi$).
\par
In practice, one resorts to the semiclassical approximation.
This has been attempted already in Ref.~\cite{tomimatsu}, but
here we want to employ the more systematic Born-Oppenheimer
approach (see, for instance, Ref.~\cite{bv,bfv}).
First of all, we assume that the complete wavefunction factorizes
into ``gravitational'' ($\psi$) and ``matter'' ($\chi$) parts,
\be
\Psi=\psi(m)\,\chi(\phi;m)
\ ,
\label{fac}
\ee
from which we get, upon substituting into Eq.~(\ref{wdw}), the
coupled equations
\be
\left\{\begin{array}{l}
\hat\Pi_m\,\tilde\psi=\expec{\hat H_\phi}\,\tilde\psi
\\
\\
\hat\Pi_m\,\tilde\chi=\left[\hat H_\phi-
\expec{\hat H_\phi}\right]\,\tilde\chi
\ ,
\end{array}\right.
\label{eqs}
\ee
where we have introduced the rescaled functions
\be
\tilde\psi=\psi\,e^{+{i}\,\int\expec{\hat\Pi_m}\,dm}
\ ,
\ \ \ \
\tilde\chi=\chi\,e^{-{i}\,\int\expec{\hat\Pi_m}\,dm}
\ ,
\ee
and the scalar field Hamiltonian
\be
H_\phi={1\over 2}\,\left[{\Pi_\phi^2\over 4\,m^2}+{1\over 2}
\right]
\ .
\label{Hph}
\ee
In the above
$\expec{\hat A(m)}\equiv\int d\phi\,\chi^*(\phi;m)\,\hat A\,
\chi(\phi;m)$ for any operator $\hat A$.
It is noticeable that Eqs.~(\ref{eqs}) do not contain terms
corresponding to gravitational fluctuations of the form which is
obtained when the momentum $\Pi_m$ enters quadratically
\cite{bfv,ebo}.
This signals the fact that, on restricting the support
of the constraints just on the apparent horizon, one is actually
freezing most of the quantum fluctuations for the system
\cite{englert}.
\par
We now assume that the variable $m$ behaves (almost) classically.
Correspondingly, the gravitational wave function $\tilde\psi$
is a wave-packet peaked on a classical trajectory $m=m(t)$,
thus it is well approximated by the WKB form
\be
\tilde\psi\simeq
{e^{i\,S_{cl}[m]}\over \sqrt{\partial_m S_{cl}}}
\ ,
\ee
where the classical action $S_{cl}$ in the exponent is evaluated
along $m(t)$.
The time variable $t$ can then be naturally related to the mass
$m$ by
\be
{\partial\over\partial t}\equiv
2\,\left[i\,{\partial\over\partial m}(\ln\tilde\psi)
-{1\over 4}\right]\,{\partial\over\partial m}
=\dot m\,{\partial\over\partial m}
\ .
\label{time}
\ee
To leading order (in $\ell_p\,m_p=\hbar$) the first of
Eqs.~(\ref{eqs}) then gives the Hamilton-Jacobi equation
\be
\dot m=2\,\left[\expec{\hat H_\phi}-{1\over 4}\right]
={\expec{\hat\Pi_\phi^2}\over 4\,m^2}
\ ,
\label{semi}
\ee
where we used $\partial_m S_{cl}=\Pi_m$, and the second of
Eqs.~(\ref{eqs}) is unaffected by the above approximation.
If $\expec{\hat\Pi_\phi^2}=k^2$ and constant, we recover the result
of Ref.~\cite{tomimatsu}.
In particular, for $k=i\,\kappa_2$ one has
\be
\dot m=-{\kappa^2_2\over 4\,m^2}
\ ,
\label{standard}
\ee
which is the well known Hawking's law of black hole decay
\cite{hawking} implying a finite time of evaporation (as measured
by a static observer).
The numerical coefficient $\kappa_2$ is arbitrary and its value could
possibly be computed by considering the constraints over the complete
space-time manifold.
\par
Instead of assuming that the state of the scalar field remains a
(momentum) eigenstate, we can use the second of Eqs.~(\ref{eqs})
to determine the evolution of the state of $\phi$, thus including
the backreaction of the change in mass onto the radiation.
Upon introducing the time according to Eq.~(\ref{time}) and making
use of Eq.~(\ref{semi}), we get the Schr\"odinger-like equation
\be
i\,{\partial\chi_s\over\partial t}
=-{\dot m\over 8\,m^2}\,\hat\Pi_\phi^2\,\chi_s
\equiv\hat{\mathcal H}_\phi\,\chi_s
\ ,
\label{sch}
\ee
where $\chi_s=\tilde\chi\,e^{-i\,\int\expec{\hat{\mathcal H}_\phi}\,dt}$.
\par
Schr\"odinger equations with time-dependent parameters can be
solved by introducing invariant operators \cite{lewis}, whose form
is explicitly known for the generalized harmonic oscillator
(see Ref.~\cite{gao} for the details).
In order to apply this method to Eq.~(\ref{sch}) with an arbitrary
$m=m(t)$, one must first ``regularize'' the Hamiltonian
${\mathcal H}_\phi$ by generating a non-vanishing
``oscillator frequency''.
This can be done by introducing a (small) harmonic-oscillator-like
potential~\footnote{The $\epsilon$-term can be viewed as a small mass
for the scalar particle (IR cut-off) or as a prescription
to discretize the spectrum.},
\be
{\mathcal H}_\phi\to{\mathcal H}_\phi+{\epsilon\over 2}\,\phi^2
\ ,
\ee
where $\epsilon\to 0$ at the end of the computation.
One then finds the spectrum $\{\ket{n,t}\,|\, n\in\natural\}$ of (exact)
solutions to Eq.~(\ref{sch}) and, on using Eq.~(\ref{semi}),
\be
\dot m^2&=&2\,\bra{n,t}\,\hat {\mathcal H}_\phi\,\ket{n,t}
\nonumber\\
&=&\left(n+{1\over 2}\right)\,\left[
\epsilon\,x^2-4\,m^2\,{\dot x^2\over\dot m}
-{\dot m\over 4\,m^2\,x^2}\right]
\ ,
\label{dotM}
\ee
where the auxiliary function $x=x(t)$ must satisfy
\be
4\,{d\over dt}\left(m^2\,{\dot x\over\dot m}\right)
-\epsilon\,x={\dot m\over 4\,m^2\,x^3}
\ ,
\label{pinney}
\ee
with suitable initial conditions.
The latter must be derived from some physical argument, as we shall
attempt in the following.
We further note that, for an evaporating black hole, $\dot m<0$
and the right hand side of Eq.~(\ref{dotM}) is consistently
positive.
\par
We shall assume that, for large mass ($m\gg 1$), Hawking's
behaviour (\ref{standard}) be recovered and determine
the corresponding $x=x_H(t)$ by expanding in powers of $1/m$,
\be
x_H=\sum_{i\ge 0}{x_i\over m^i}
\ ,
\label{SxH}
\ee
where the coefficients $x_i=x_i(t)$.
Inserting the sum (\ref{SxH}) into Eqs.~(\ref{dotM}) and (\ref{pinney})
yields, in the limit for $\epsilon\to 0$,
\be
x_H=\sqrt{{2\,n+1}\over 2\,\kappa_2}+{\mathcal O}\left({1\over m^8}\right)
\ ,
\ee
and one can further check that Eq.~(\ref{standard}) is a consistent
solution to the system of Eqs.~(\ref{dotM}) and (\ref{pinney}) to even
higher orders (details will be given elsewhere \cite{next}).
\par
\begin{figure}
\centering
\raisebox{5cm}{$\dot m$}\hspace{0cm}
\epsfxsize=3.5in
\epsfbox{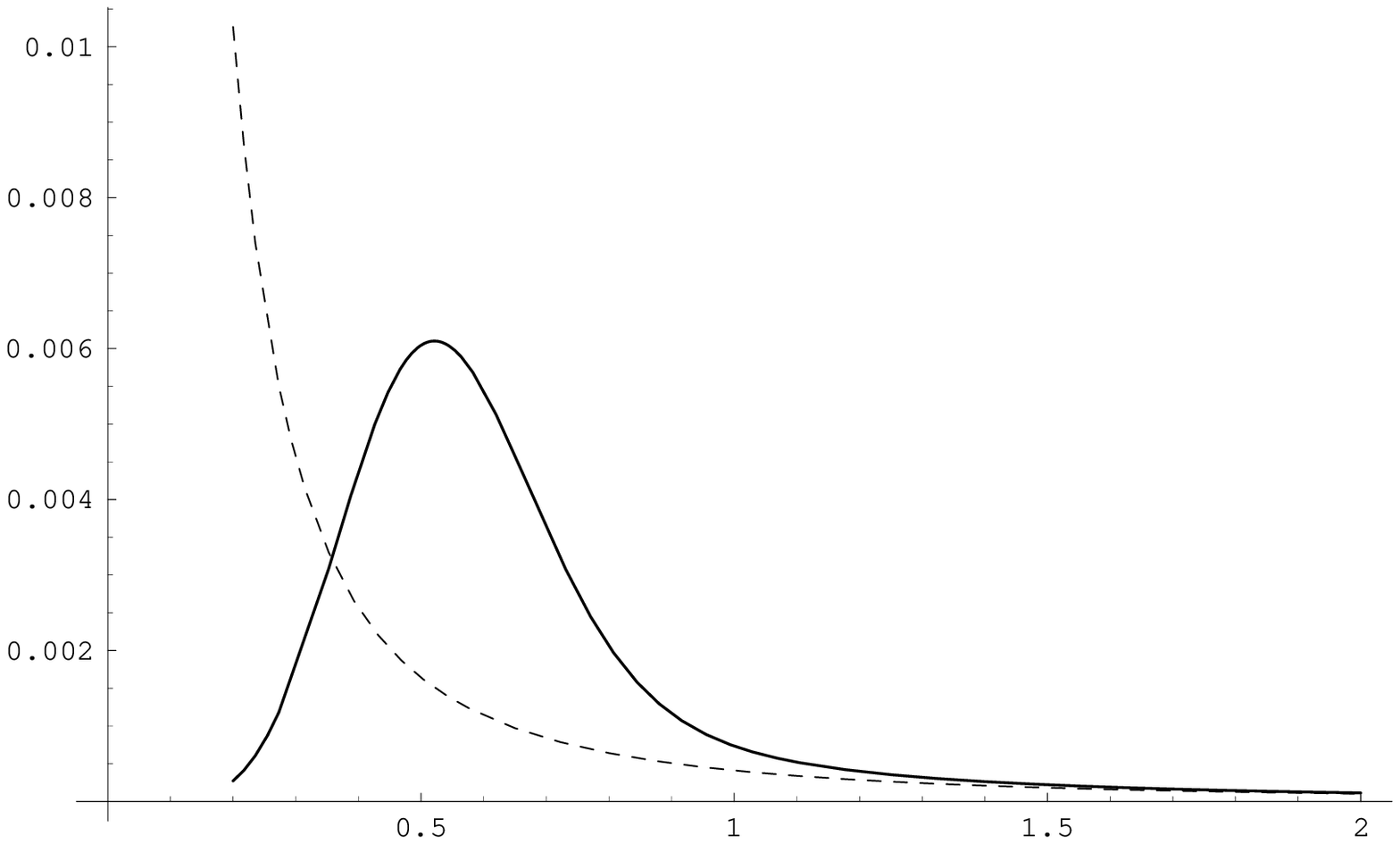}\\
\raisebox{0.5cm}{\hspace{7cm} $m$}
\caption{Luminosity of a radiating black hole in the canonical
(dashed line) and microcanonical (continuous line) pictures.}
\label{microl}
\end{figure}
\begin{figure}
\centering
\raisebox{5cm}{$m$}\hspace{0cm}
\epsfxsize=3.5in
\epsfbox{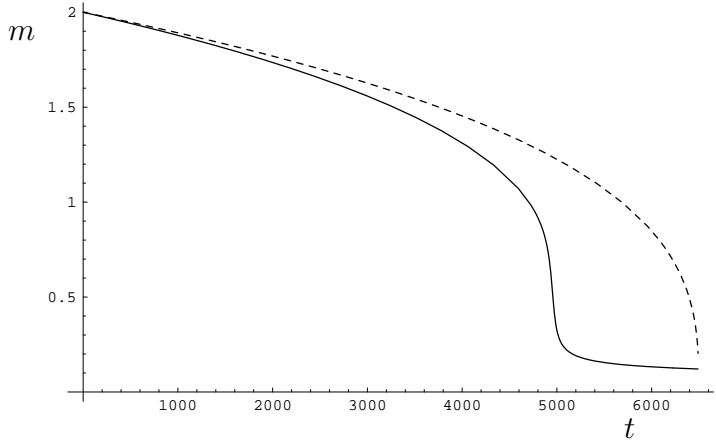}\\
\raisebox{0.5cm}{\hspace{7cm} $t$}
\caption{Time evolution of the radiating black hole in the
canonical (dashed line) and microcanonical (continuous line) pictures.}
\label{microt}
\end{figure}

From the microcanonical description we know that the law
(\ref{standard}) must be corrected for small masses.
In fact the correct number density of Hawking quanta is better
approximated by the microcanonical expression \cite{micro}
\be
N(\omega)=\sum_{l=1}^{[[m/\omega]]}\,
{\exp\left[4\,\pi\,(m-l\,\omega)^2\right]
\over{\exp(4\,\pi\,m^2)}}
\ ,
\ee
where $[[X]]$ denotes the integer part of $X$.
In the limit $m \to \infty$, $N$ equals the canonical ensemble
number density (Planckian distribution).
The decay rate (the opposite of the luminosity) for a black hole
is then given by
\be
\dot m=-{\mathcal A}\,\sum_k\,\int_0^{\infty}
N_k\,(\omega)\,\Gamma_k(\omega)\,\omega^3\,d\omega
\ ,
\label{dMdt}
\ee
where $\Gamma$ is the grey-body factor and the sum is over
particle species and angular momentum \cite{macgibbon}.
The comparisons between the luminosity and time evolution of $m$
in the canonical and microcanonical pictures are displayed,
respectively, in Fig.~\ref{microl} and Fig.~\ref{microt} for
the simplified case $\Gamma=k=1$.
The basic feature of the microcanonical evolution is the existence
of a long tail due to the suppression of (relatively) high energy
modes.
\par
Bearing in mind the above result, we can try and perturb Eq.~(\ref{standard})
by assuming an expansion in powers of $1/m$,
\be
&&\dot m=-{\kappa_2\over 4\,m^2}+\sum_{i\ge 3}{\kappa_i\over m^i}
\nonumber \\
\\
&&x=x_H+\sum_{i\ge 1}{y_i\over m^i}
\ ,
\nonumber
\ee
where $\kappa_i=\kappa_i(t)$ and $y_i=y_i(t)$.
It turns then out that Eqs.~(\ref{dotM}) and (\ref{pinney})
admit as solutions
\be
\dot m=-{\kappa_2\over 4\,m^2}+{\kappa_a\over m^a}
+{\mathcal O}\left({1\over m^{a+1}}\right)
\ ,
\label{non-standard}
\ee
with $a\ge 3$, $\kappa_a$ constant, and a corresponding function
$x=x(t)$ (whose form is of no relevance here \cite{next}).
Again the value of the coefficients in the above equation cannot
be fixed within the dynamical approach.
\par
\begin{figure}
\centering
\raisebox{5cm}{$m$}\hspace{0cm}
\epsfxsize=3.5in
\epsfbox{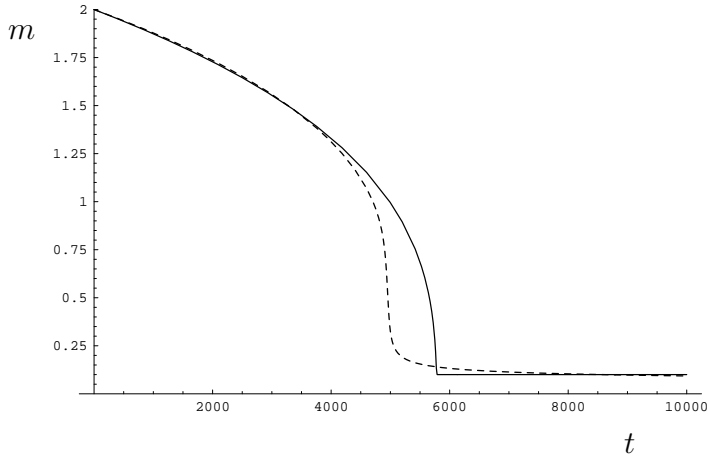}\\
\hspace{7cm} $t$
\caption{Time evolution of the radiating black hole from the Hamiltonian
constraint (continuous line) as compared with the microcanonical picture
(dashed line) for the case described in the text.}
\label{decay}
\end{figure}

One can fix the values of $\kappa_2$ and $\kappa_a$ in such a way that
the luminosity given by Eq.~(\ref{non-standard}) approximates
the microcanonical luminosity of Fig.~\ref{microl} and the history
of the mass approximates that obtained from the microcanical
picture, at least for $m>1$.
In Fig.~\ref{decay} we plot $m=m(t)$ for $\kappa_6=10^{-4}\,\kappa_2$
and compare it with the microcanonical history.
Since we used the large $m$ approximation in order to solve
Eqs.~(\ref{dotM}) and (\ref{pinney}), the plot is not reliable for
$m<1$, although it is certainly suggestive that both curves show the
presence of a long-living remnant of sub-Planckian mass.
\par
We wish to conclude by remarking that the main result of this
letter is given by the set of coupled Eqs.~(\ref{dotM}) and
(\ref{pinney}) which govern the dynamics of an evaporating
(apparent) horizon and allow one to study the backreaction
of the emitted radiation.
Unfortunately, they are rather complicated and do not allow
for a straightforward analytical treatment.
Further, one needs to specify the initial conditions in such a
way that relevant physical situations are described.
This far, we have just investigated that system of equations
by employing the large mass approximation and compared with the
results obtained from the statistical mechanics of the Hawking
radiation in order to include the backreaction.
We hope to carry on a more complete (possibly numerical) analysis
and gain a better understanding of the initial conditions in a
forthcoming work \cite{next}.

\begin{thebibliography}{99}
%
\bibitem{hajicek}
P. Hajiceck, Nucl. Phys. {\bf B} (Proc. Suppl.) (2000) {\bf 88}
114.
%
\bibitem{hawking}
S.W. Hawking, Nature {\bf 248} (1974) 30 and
Comm. Math. Phys. {\bf 43} (1975) 199.
%
\bibitem{dewitt1}
B. S. DeWitt, in {\em Relativity, groups and topology}, Vol.~1,
DeWitt C. and DeWitt B. S. editors (Gordon and Breach, New York,
1964).
%
\bibitem{birrell}
N. D. Birrell and P.C.W. Davis, {\em Quantum fields in curved
space} (Cambridge, Cambridge University Press, 1982).
%
\bibitem{dewitt}
B. S. DeWitt, Phys. Rev. {\bf 160} (1967) 1113.
%
\bibitem{wheeler}
J. A. Wheeler, in {\em Batelle rencontres: 1967 lectures in
mathematics and physics}, C. DeWitt and J. A. Wheeler editors
(Benjamin, New York, 1968).
%
\bibitem{bv}
R. Brout and G. Venturi, Phys. Rev. D {\bf 39} (1989) 2436.
%
\bibitem{lousto}
C.O. Lousto and F.D. Mazzitelli, Int. J. Mod. Phys.
{\bf A 6} (1991) 1017.
%
\bibitem{tomimatsu}
A. Tomimatsu, Phys. Lett. {\bf B289} (1992) 283.
%
\bibitem{bekenstein}
J.D. Bekenstein, Phys. Rev. D {\bf 7} (1973) 2333;
Phys. Rev. D {\bf 9} (1974) 3292.
%
\bibitem{micro}
B. Harms and Y. Leblanc, Phys. Rev. D {\bf 46} (1992) 2334;
Phys. Rev. D {\bf 47} (1993) 2438;
R. Casadio and B. Harms, Phys. Rev. D {\bf 58} (1998) 044014;
Mod. Phys. Lett. {\bf A17} (1999) 1089.
%
\bibitem{strings}
A. Strominger and C. Vafa, Phys. Lett. {\bf B379} (1996) 99;
J. Maldacena, {\em Black holes in string theory}, hep-th/9607235.
%
\bibitem{page}
D. Page, Phys. Rev. D {\bf 13} (1976) 198;
Phys. Rev. D {\bf 16} (1977) 2401.
%
\bibitem{thomi}
P. Thomi, B. Isaak and P. Hajicek, Phys. Rev. D {\bf 30}
(1984) 1168;
P. Hajicek, Phys. Rev. D {\bf 30} (1984) 1178.
%
\bibitem{mtw}
C.W. Misner, K.S. Thorne and J.A. Wheeler,
{\em Gravitation} (San Francesco: Freeman, 1973).
%
\bibitem{bfv}
C. Bertoni, F. Finelli and G. Venturi, Class. Quantum Grav.
{\bf 13} (1996) 2375.
%
\bibitem{ebo}
R. Casadio, Int. Jour. Mod. Phys. D {\bf 9} (2000) 511.
%
\bibitem{englert}
F. Englert, Phys. Lett. {\bf B 228} (1989) 111.
%
\bibitem{lewis}
H.R. Lewis and W.B. Riesenfeld, J. Math. Phys. {\bf 10}
(1969) 1458.
%
\bibitem{gao}
X.C. Gao, J.B. Xu and T.Z. Quian, Phys. Rev. A {\bf 44}
(1991) 7016.
%
\bibitem{next}
R. Casadio, in preparation.
%
\bibitem{macgibbon}
F. Halzen, E. Zas, J.H. MacGibbon and T.C. Weeks, Nature {\bf 353}
(1991) 807.
%
\end{thebibliography}
\end{document}